\documentclass[aps,prl,twocolumn,showpacs,superscriptaddress,groupedaddress]{revtex4-1}

\usepackage{mathrsfs}
\usepackage{graphicx}
\usepackage{amsmath}
\usepackage{epstopdf}
\usepackage{color}
\usepackage{soul}
\def\red{\textcolor{red}}
\def\blue{\textcolor{blue}}
\definecolor{darkblue}{RGB}{83,0,93}

\newsavebox{\astrutbox}
\sbox{\astrutbox}{\rule[-5pt]{0pt}{20pt}}

\def\be{\begin{eqnarray}}
\def\ee{\end{eqnarray}}
\def\bes{\begin{subeqnarray}}
\def\ees{\end{subeqnarray}}

\def\f{\frac}
\def\lp{\left(}
\def\rp{\right)}
\def\lb{\left[}
\def\rb{\right]}

\def\n{\nabla}

\def\befi{\begin{figure}}
\def\eefi{\end{figure}}

\def\bce{\begin{center}}
\def\ece{\end{center}}

\def\L{\Lambda}

\def\n{\nabla}

\def\i{\textrm{i}}

\def\ba#1\ea{\begin{align}#1\end{align}}

\def\bsa#1\esa{\begin{subequations}
\begin{align}#1\end{align} \end{subequations}}

\usepackage{graphicx}
\usepackage{dcolumn}
\usepackage{bm}
\def\blue{\textcolor{blue}}                                                     
\def\red{\textcolor{red}}                                                       
                                                   
\def\black{\textcolor{black}}                                                   
                                                     
\definecolor{darkblue}{RGB}{83,0,93}                                            
\def\L{\mbox{------}}                                                           
\def\dashL{\mbox{-~-~-}}                                                        
                                         
\def\cdashL{\mbox{---~-~---}}

\def\blue{\textcolor{blue}}                                                     
\def\red{\textcolor{red}}                                                       
                                                   
\def\black{\textcolor{black}}                                                   
  
\begin{document}

\preprint{APS/123-QED}

\title{Broadband Wall-less Waveguide for Shallow Water
  Waves}

\author{Ahmad Zareei} \email{zareei@berkeley.edu}
\altaffiliation{Mechanical Engineering Department, University of
  California,
  Berkeley 94720, USA}
\author{Mohammad-Reza Alam}%
\affiliation{Mechanical Engineering Department, University of
  California,
  Berkeley 94720, USA}




\begin{abstract}


  We present a broadband waveguide for water waves obtained through
  mere manipulation of seabed properties and without any need for
  sidewalls. Specifically, we show that a viscoelastic seabed results
  in a modified effective gravity term in the governing equations of
  water waves, which provides a generic broadband mechanism to control
  oceanic wave energy and enables confining surface waves inside a long
  narrow path without sidewalls. Our findings have promising
  applications in guiding and steering waves for oceanic wave energy
  farms or protecting shorelines.

\end{abstract}

\maketitle


\section{Introduction}

The Intensity of all physical waves decreases as they propagate away
from their sources due to the so-called \textit{spreading
  loss}. Waveguides counteract this spreading and keep the energy
flowing over long distances with minimal losses.  The simplest
waveguide is based on reflection of waves from waveguide's boundary
due to a sharp change in refractive index
\cite{thomson1893,rayleigh1897,rayleigh1896}. For electromagnetic
waves, optical fibers (or dielectric slabs) are an example of such
waveguides where light rays entering the fiber with incidence angle
upto the fiber's cut-off angle \cite{jenkins1937fundamentals} get
reflected from boundaries, stay confined, and propagate inside the
fiber \cite[][]{keiser2003optical}. An equivalent of such simple
waveguide for water waves is a side-wall confinement where water waves
propagate inside a \textit{canal} while getting reflected from
boundaries. Additionally, the propagation dynamics of various surface
gravity water wave envelopes in canals has been extensively studied
\cite[][]{fu2015propagationa, fu2015propagationb, fu2017dispersion}.


The drawback of reflection-based waveguides is modal dispersion
which effectively spreads the temporal duration of a pulse and limits
the operational bandwidth \cite{pollock2003integrated}.
Modal dispersion may considerably be reduced by engineering the
velocity profile and therefore the refractive index distribution
inside a waveguide. A graded core fiber with a parabolic refractive
index profile \cite{adams1981introduction} is an example of such
waveguides for electromagnetic waves, where wave rays refract toward
the fiber's centerline as they travel through the waveguide
\cite[][]{feit1978light,zareei2018,Wolfe12434,ansell2015hybrid}. Shallow
water waves behave similar to optical waves, where water wavenumber
$k$ is equivalent to optical refractive index $n$ and it satisfies the
Fermat's principle similar to optics \cite{Mei2005}. The wavenumber in
surface gravity waves depends on the water depth and gravitational
acceleration through $c^2 = gh$, where $c=\omega/k$ is the phase
speed, $g$ is the gravitational acceleration, and $h$ is the water
depth. As a result, any variation in water depth or gravitational
acceleration alters the refractive index profile which would enable
wave ray path engineering in water waves, e.g., a ridge-form sea-bed
topography has a refractive index similar to a graded-index fiber and
is capable of trapping certain frequencies  of long-wave
oceanic waves over the ridge
\cite[][]{buchwald1969long,mei1989theory}. Besides refractive index,
the effects of depth variation on water wave propagation can be
considered through other mechanisms as well, e.g., through (i) Bragg
resonance between surface waves and sea-bed topography
\cite{Liu1998,ALAM2010,Elandt2014}; (ii) band features of periodic
structures inspired by photonic crystals \cite{Hu2003, Hu2003a,Hu2004,
  Jeong2004,Shen2005,Shen2005a,Bobinski2015}; (iii) homogenization
effect of a rapidly changing topography
\cite{Berraquero2013,maurel2017revisiting,Zareei2015a,zareei2017}. The first two
methods are narrow band and highly sensitive to frequency, while the
last approach works in a broader range of frequencies. The effective
bandwidth of last approach, the homogenization method, can be found by
homogenization condition which restricts the incident wavelength to be
much larger than characteristic length of sea-bed
variations.



The other parameter affecting water waves, in addition to water depth,
is the gravitational acceleration which on the contrary is always a
physical constant and can not be altered. While the actual gravity is
constant, the observed effective gravity can change. There are only a
few studies observing an altered effective gravitational acceleration
for water waves: (i) an array of vertical cylinders at low frequencies
simultaneously changes both effective depth and effective gravity
which depends on the filling ratio of cylinders \cite{Hu2005}; (ii) a
periodic array of resonators at low frequencies has a negative
effective gravity which results in band-gap that strongly reflects the
waves \cite{Hu2011,Hu2013}; (iii) a thick, rigid, and unmovable
plate covering the water surface exhibits an infinite gravitational
acceleration (or equivalently a zero refractive index) which is
analogous to epsilon near zero materials in optics
\cite{edwards2008experimental,maas2013experimental,moitra2013realization}
and has potential applications in focusing and collimation of water
waves \cite{Zhang2014}. Although these methods offer scenarios in
which an altered effective gravitational acceleration is observed
(only infinite or negative); they do not provide a generic mechanism
to obtain finite variations in effective gravitational
acceleration. It is to be noted that arbitrary modification of
effective gravity enables \textit{broadband} manipulation of shallow
water waves which, on the contrary, is not possible through water
depth variations. Furthermore, the transformation optics method
usually requires variations in the gravitational acceleration
\cite[][]{Zareei2015a} which enforces using alternative methods
\cite[e.g.,][]{li2018concentrators,zareei2016cloaking}.

Here, we show that a visco-elastic bottom topography results in an
effective gravitational acceleration which can be adjusted through
elasticity and damping of a sea-bed carpet. The tunable effective
gravitational acceleration
provides a generic mechanism to control the flow of ocean wave energy
through a passive sea-bed, and is exemplified by a water waveguide
which has no physical walls and still confines the water waves to
propagate inside the waveguide region. The water waveguide is
analogous to graded index optical fibers with parabolic refractive
index and is engineered through the effective gravitational
acceleration using elastic sea-bed carpet. Water waves over the
designed elastic sea-bed (waveguide) are transmitted over long
distances with a low attenuation loss. We numerically validate the
design and test the effectiveness of the waveguide in transmitting
water wave energy.
The proposed mechanism along with water depth variations
open new venues in the control of ocean wave energy which have
engineering applications in guiding and steering incoming waves toward
or away from destinations of interest such as artificial surf zones,
wave energy farms, or to protect shorelines.



\section{Governing Equations}

\begin{figure}[!h]
  \centering
  \includegraphics[width=3.3in]{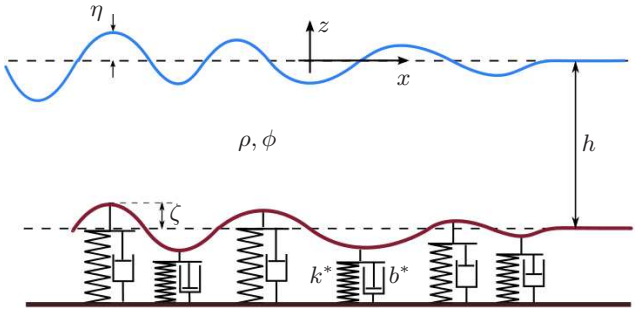}
  \put(-206,111){$\eta$}
  \put(-120,105){$z$}
  \put(-90,85){$x$}
  \put(-20,60){$h$}
  \put(-150,62){$\rho,\phi$}  
  \put(-176,34){$\zeta$}
  \put(-93,10){$b^{*}$}
  \put(-123,10){$k^{*}$}
  \caption{Schematic of a visco-elastic carpet on the seafloor. The
    carpet is composed of linear springs with stiffness coefficient
    $k^{*}$ providing the restoring force, and dash-pot type dampers
    with damping coefficient $b^{*}$ extracting energy from the
    system. The distance between each module of the spring damper is
    assumed to be much smaller than the typical wavelength of the over
    passing waves such that the assumption of continuously distributed
    springs and dampers is valid.}\label{fig100}
\end{figure}
  We consider an inviscid, incompressible, and irrotational fluid over a
visco-elastic sea-bed (see Fig. \ref{fig100}). We set the coordinate system on the mean free
surface of water with $z$-axis pointing upward and the mean depth being at
$z=-h$. The linearized governing equations for the surface/bottom elevation
$\eta/\zeta$ and the velocity potential $\phi$, ignoring the surface
tension, are \cite{Alam2012b}
\begin{subequations}\label{eq1}
    \begin{align}
      &\n^2 \phi =0, \qquad -h \leq z \leq 0 \label{eq1a} \\
      & \eta_t  =  \phi_z  , \qquad z= 0 \label{eq1b}\\
      &\phi_t  + g\eta = 0 , \qquad z =0 \label{eq1c} \\
      &\zeta_t - \lp h_x \phi_x + h_y\phi_y\rp = \phi_z, \quad z = -h  \label{eq1d} \\
      &\phi_t  + g\zeta + \f{1}{\rho} \lb b^*  \zeta_t + k^* \zeta \rb = 0 , \quad z =-h \label{eq1e}
    \end{align}
\end{subequations}
where $k^*$ and $b^*$ are the stiffness and viscous damping
coefficient of the visco-elastic bottom per unit area, $\rho$ is the
density of fluid, and $g$ is the gravitational acceleration. In this
set of governing equations, \eqref{eq1a} is the continuity equation in
the fluid domain, \eqref{eq1b} and \eqref{eq1d} are the kinematic
boundary conditions on the free surface and bottom topography, and
\eqref{eq1c} and \eqref{eq1e} are dynamic boundary conditions on the
free surface and visco-elactic bottom. We nondimensionalize the
governing equations with the following nondimensional groups as
\begin{subequations}
\begin{align}
  &(x',y') = k(x,y), \quad z' = \f{z}{H}, \quad h' = \f{h}{H}, \quad t' = k\sqrt{gH} t,\\
  & \eta' = \f{\eta}{A}, \quad \phi' = \lp \f{1}{k} \f{A}{H} \sqrt{gH}\rp^{-1} \phi \\
&\zeta' = \f{\zeta}{A}, \quad \lambda = (kH) \f{b^*}{\rho \sqrt{gH}}, \quad \gamma = \f{k^*}{\rho g}, \quad \mu \equiv kh
\end{align}
\end{subequations}
where $k$ and $A$ are respectively the characteristic wave number and wave
amplitude of incident waves. Nondimensionalizing the
governing equations in (\ref{eq1}), dropping the primes for the sake
of simplicity, and assuming of time harmonicity $\omega$ for the
incoming waves, the governing equations simplify to
%
%
\begin{subequations}
\begin{align}
  &\mu^2 \lp \phi_{xx} + \phi_{yy} \rp + \phi_{zz} = 0, -h\leq z  \leq 0\\
  &- \mu^2 \omega^2 \phi +  \phi_z = 0, \qquad z =0 \\
  & \mu^2 \f{ \omega^2  \phi}{1 +  \lambda (\i \omega)   + \gamma } -\mu^2 \lp h_x \phi_x + h_y \phi_y \rp = \phi_z, \qquad z = -h  .
\end{align}
\end{subequations}
Next, we assume the shallow water regime, $\mu \ll 1$, and expand the
solution for the velocity potential as
$\phi = \phi^{(0)} + \mu^2 \phi^{(2)} + \ldots$. Finding the zeroth
and second order solutions, and considering the solvability condition
\cite{Mei2005}, one can obtain the zeroth mode governing equation. In
the dimensional form, the governing equation reads
\begin{align}\label{eq555}
  \n. \lp h \n \phi^{(0)}  \rp + \f{\omega^2}{\tilde g}  \phi^{(0)} =0, \qquad   \tilde g= g\lb 1 + \f{\rho g}{k^* + \i b^*\omega}\rb.
\end{align}
The obtained governing equation \eqref{eq555} is similar to the shallow water wave
equation \cite{Mei2005}; however, with a different gravitational
acceleration $\tilde g$. This derivation suggests that the net effect
of a visco-elastic sea-bed in shallow water regime can be summarized
with a modified gravity term that depends on the elasticity and
viscosity of the sea-bed.

\section{Broadband Waveguide Design}

In an incompressible, inviscid, and irrotational fluid where the fluid
depth is $h$, the wave equation in long wave limit $kh\ll 1$ reads as
$\n . (h \n \phi) + \omega^{2}\phi/g = 0$, where $\nabla$ is the
horizontal gradient, $\phi$ is the velocity potential with fluid
velocity $\mathbf{v} = \n \phi$, $\omega$ is the wave frequency, and
$g$ is the gravitational acceleration. Equivalent refractive index for
water waves is obtained as $n = c_0/c$, where $c_0$ is the wave phase
velocity at a reference point and $c$ is the local phase velocity
\cite{Mei2005}. Since phase velocity is inversely proportional to the
wave number, $c=\omega/k$, the refractive index becomes proportional
to the wave number as $n\propto k$. Phase velocity of shallow water
waves depends on the water depth and gravitational acceleration
through $c = \sqrt{gh}$, where variations in water depth $h$, or
gravitational acceleration $g$ would lead to changes in the refractive
index $n$. The local refractive index determines the wave ray paths
for water waves, and therefore enables steering of water waves through
refractive index profile engineering.  In shallow water waves, water
depth is constrained by $kh\ll 1$, and as a result water depth
variation decreases the effective bandwidth of the system. The only
other parameter that effects the refractive index is the gravitational
acceleration which is a physical constant. Nevertheless, a
visco-elastic sea-bed results in an effective (modified) gravitational
acceleration which can be adjusted by changing the elasticity and
viscosity of the sea-bed. The modified gravitational acceleration
allows variations in refractive index and contrary to water depth
variations does not restrict the effective bandwidth of the system.

Assuming a visco-elastic sea-bed, the governing equation for shallow water waves is
obtained as $\n. (h \n \phi) + \omega^{2}\phi/\tilde g = 0$, where
\begin{align}\label{eq100}
  \tilde g = g \lb 1 + \frac{\rho g}{k^{*} + \i b^{*} \omega} \rb,
\end{align}
in which $k^{*},b^{*}$ are the spring and damping coefficient of the
visco-elastic bottom, $\rho$ is the fluids density, and $\omega$ is
the frequency of incoming wave. Therefore,
the effective gravitational acceleration observed is different from
constant gravitational acceleration $g$ and can be altered using
stiffness and damper coefficients in the visco-elastic sea-bed. In the
limit where the springs are stiff $k^{*}/\rho g \gg 1$, the effective
gravitational acceleration simplifies to $\tilde g = g$ as
expected. The imaginary part of the effective gravitational
acceleration depends on both frequency and damper coefficient
$\i b^{*}\omega$ and represents the absorption of energy by the
dampers in the visco-elastic
carpet and can be used in designing high-performance wave energy
extraction devices through actuated sea-floor mounted carpets \cite{desmars2018interaction,Alam2012b}.

Using obtained effective gravity in \eqref{eq100}, the refractive index
becomes
\begin{align}\label{eq200}
  n = \frac{c_{0}}{\sqrt{\tilde g h}}  = \sqrt{\frac{h_0}{h}} \frac{1}{\sqrt{ 1 + {\rho g}/\lp {k^{*} + \i b^{*} \omega} \rp }}, 
\end{align}
where it depends on the visco-elastic sea-bed parameters. This
provides a generic broadband mechanism to alter the refractive index
through elasticity and viscosity of the sea-bed, and hence enables
control of wave rays path of oceanic water waves.  Here as an example,
we design a visco-elastic sea-bed analogous to graded-index optical
fibers for guiding water waves.

\begin{figure}[!h]
  \includegraphics[width=3.6in]{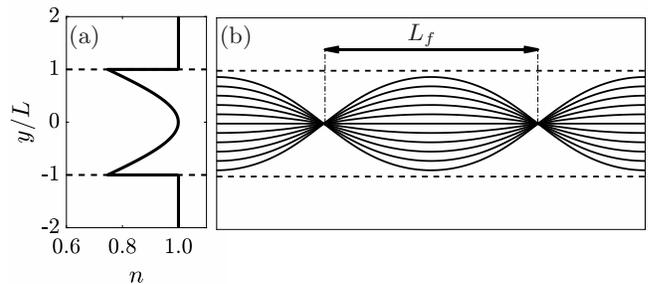}
  \put(-238,91){(a)}
  \put(-181,91){(b)}
  \put(-110,92){$L_f$}
  \caption{(a) Refractive index profile in a graded index (GRIN) wave
    guide where $n(y) = n_c \text{sech}(\kappa~y/L)$. (b) Top view of
    ray paths (solid lines) in a graded-index wave-guide (dashed
    lines). The tuning parameter $\kappa$ in refractive index profile
    controls the distance between focal points $L_f$. }\label{fig2}
\end{figure}

%
Graded-index waveguides (or fibers) in optics are used to guide
electromagnetic waves over long distances. The refractive index of
such waveguides should vary as $n(y) = n_c~\text{sech}(\kappa y/L)$,
where $n(y)$ is the refractive index of the waveguide measured by the
distance from the center line of the fiber, $n_c$ is the fiber's core
refractive index, $L$ is the width of the waveguide, and $\kappa$ is a
tuning parameter that controls the distance between focal points $L_f$ (Fig. \ref{fig2}).

\begin{figure}[!h]
  \centering
  \includegraphics[width=2.91in]{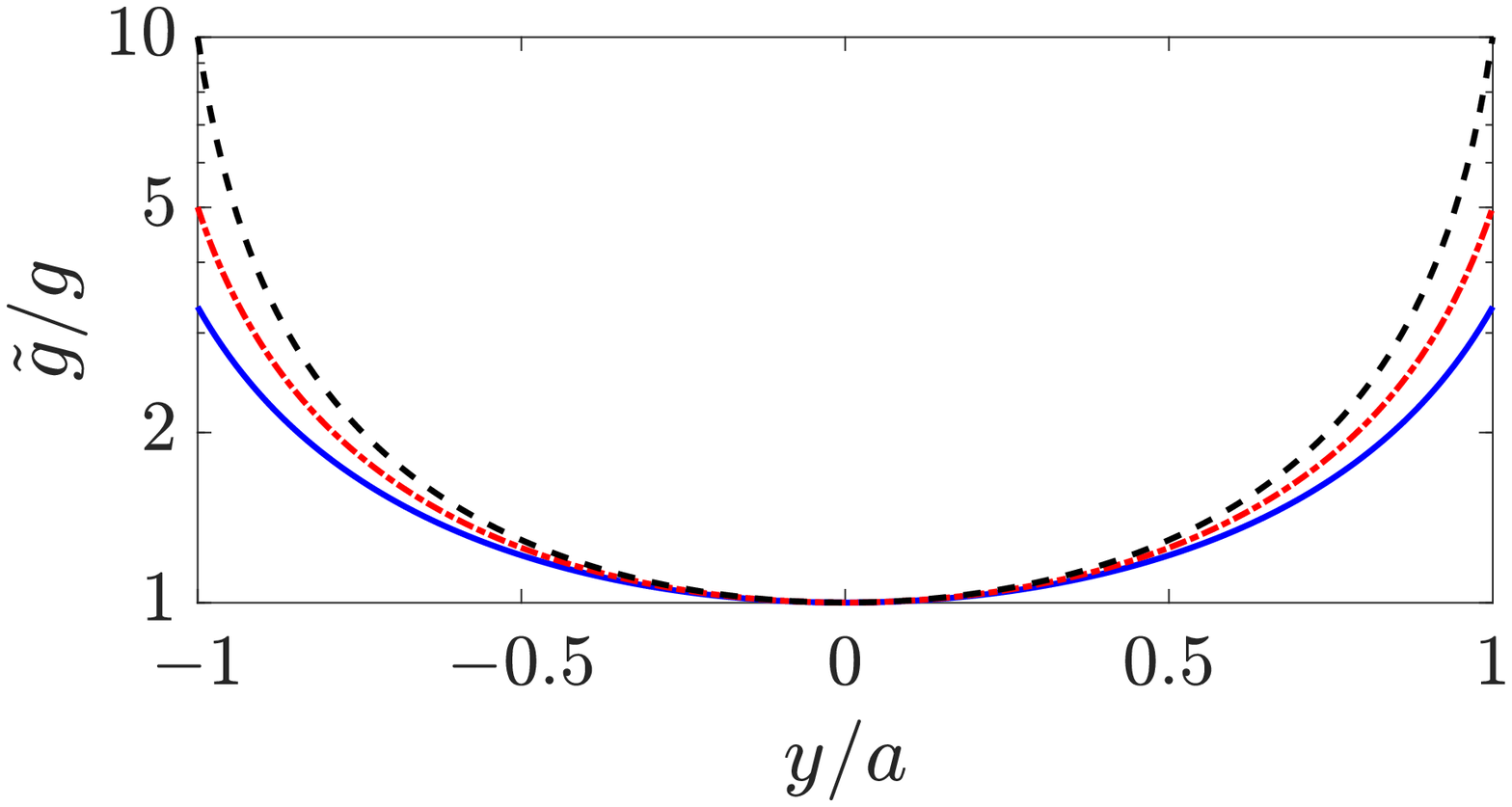} \put(-180,87){(a)}\\
  \includegraphics[width=2.91in]{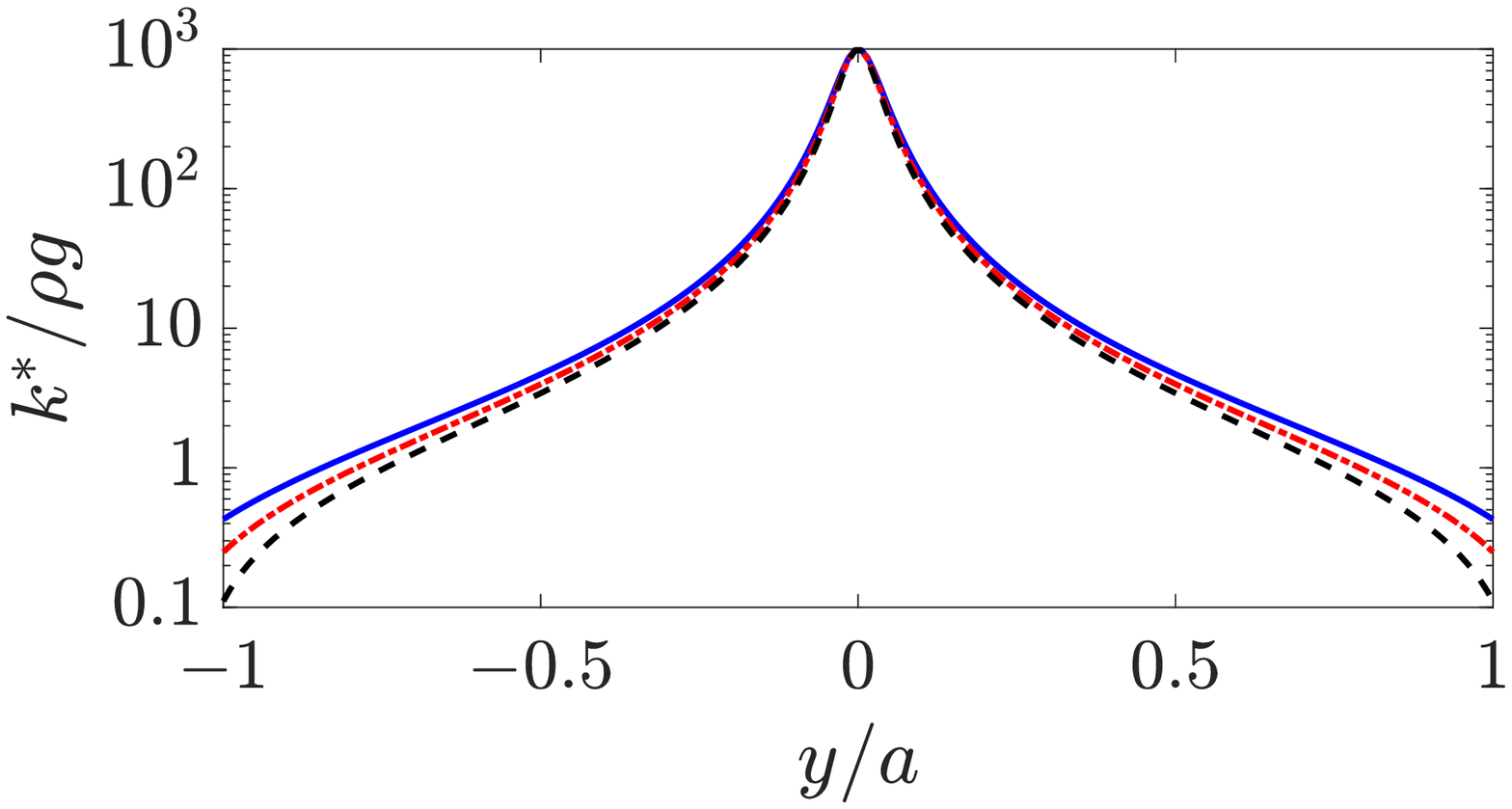}
  \put(-180,87){(b)}
  \caption{Nondimensionalized stiffness $k^{*}/\rho g$ and effective
    gravity $\tilde g/g$ over the width of the water waveguide $y/a$
    for different values of $\delta$ ($\delta=0.7$ (\blue{\L}),
    $\delta=0.8 (\red{\cdashL})$, and $\delta=0.9$ ({\dashL}))
    where the waveguide's core effective gravity is considered to
    be very close to the gravitational acceleration $\tilde g/ g = 1.001$ }
\label{fig3}
\end{figure}

For the ease of experiment, we use the first two terms in the Taylor
expansion of refractive index as \cite{Gomez2012}
\begin{align}\label{eq300}
\lp \frac{n(y)}{n_c}\rp^2 =   1 - \delta \frac{y^2}{L^2}, 
\end{align}
which gives a parabolic refractive index profile with
$\delta = \kappa^2$. Similar to optical waves, shallow water waves
stay confined inside a region with parabolic refractive index profile
and wave energy is transmitted over long distances preventing the
spreading loss. We take $x$-axis to be the center-line of the GRIN
waveguide, and $y$-axis representing the distance from waveguide's
centerline. Using equations \eqref{eq200} and \eqref{eq300}, the
effective gravity is obtained as
$g(y) = {\tilde g_{c}}/\lp {1 - \delta y^{2}/L^{2}}\rp$ where
$\tilde g_c$ represents the effective gravity at the waveguide's
centerline. The stiffness of the elastic bottom topography is then
obtained using equation \eqref{eq100} as
${\rho g}/{k^{*}} = \tilde g/g - 1 $. Figure \ref{fig3} depicts the
values of effective gravitational acceleration and elastic sea-bed
stiffness over the width of the waveguide. The profiles are plotted
for different values of $\delta$ to show the sensitivity of the
obtained profile with variations in $\delta$.  The waveguide's core
effective gravity is considered to be very close to the gravitational
acceleration $\tilde g_c/g = 1.001$ which translates into a stiff (or
rigid) sea-bed. Note that a higher difference between core's effective
gravitational acceleration and the actual gravity causes more
impedance mismatch at the entrance of the waveguide and therefore a
higher reflection at the entry of the waveguide; accordingly the
core's gravity is considered to be close to the actual gravity. In the
asymptotic, the rigid ground has stiffness of infinity where the
effective gravitational acceleration becomes the actual gravity
$\tilde g = g$.

\section{Results}

In order to validate the theoretical design, we use Finite Element
Methods (FEM) to solve the governing equation, given the stiffness
profile shown in Fig. \ref{fig3} for $\delta = 0.8$. We use
Crank-Nicolson method for time discretization along with piece-wise
linear isoparametric shape functions for spatial discretization. We
chose the open-source finite element package \texttt{FreeFEM++}
\cite{MR3043640} to solve the discretized variational formulation of
the governing equation. The length scale is non-dimensionalized using
the wave-length of the incoming wave, and the non-dimensionalized time
is obtained from dispersion relation. The numerical domain is set
to $[0,30]\times[-7.5,7.5]$, where the center of the wave guide is at
$y=0$ and spans in the $y$-direction from $-1.5$ to $1.5$.  The
domain's border is seeded by $\delta x = 0.04$ and the element mesh is
then created using \texttt{FreeFem++} mesh generator. Waves are inserted
through the waveguide using a $\tanh$ ramp-function representing the
wave generator and solution is found by marching in time with
$\delta t = T/100$ where $T$ is the wave period. Artificial absorbing
boundary condition is used to absorb the outgoing waves at the
boundaries of the domain.

%
  
%
\begin{figure}[!h]
  \centering
  \includegraphics[width=3.5in]{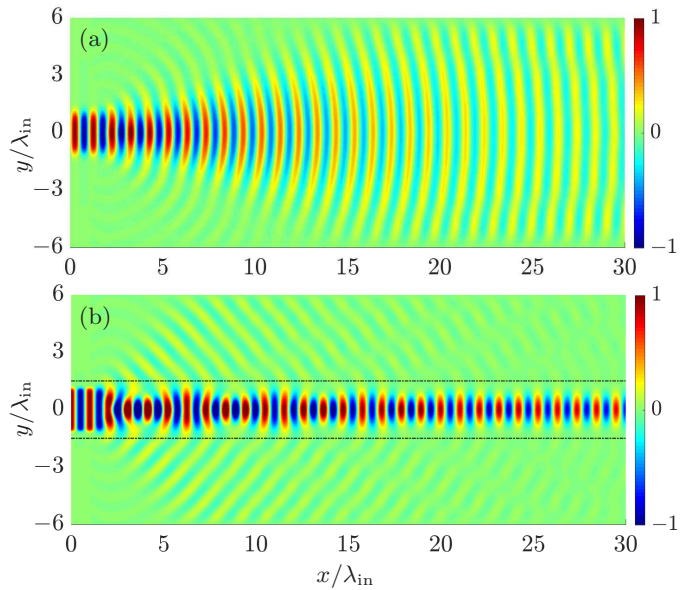}
  \put(-230,193){(a)}
  \put(-256,145){\rotatebox{90}{$y/\lambda_{\text{in}}$}}
  \put(-256,45){\rotatebox{90}{$y/\lambda_{\text{in}}$}}
  \put(-230,88){(b)}
  \put(-140,-10){$x/\lambda_{\text{in}}$}
  \caption{Top view of normalized velocity potential
    $\phi/\phi_{\text{in}}$, where $\phi_{\text{in}}$ is the incident
    velocity potential inside the test region. The normalized wave
    amplitude profile $\eta/\eta_{\text{in}}$ is the same as
    normalized velocity potential, since $\eta = \i \omega \phi / g$.
    Waves are inserted in the domain from left and propagate to
    right. Figure (a) shows the normalized velocity potential over a
    constant depth sea-bed and gravity. The waves spread as they
    propagate and the amplitude of decreases.  Figure (b) shows the
    normalized velocity potential when the elastic waveguide is
    implemented in the middle region between dashed lines
    (\black{\dashL}). The elastic sea-bed confines the waves inside
    the middle region and prevents the waves from spreading. The net effect of the elastic sea-bed is
    similar to the graded-index optical fibers that confine and
    transmit electromagnetic wave rays.}\label{fig4}
\end{figure}

As shown in Fig. \ref{fig4}a, when the bottom topography is flat,
the water waves disperse in the transverse direction normal to the
wave rays propagation and therefore the wave amplitude decreases as it
propagates through the simulation region. On the other hand, when the
graded index waveguide is implemented through elastic sea-bed, the
water waves stay confined inside the waveguide region and propagate
with much less decrease in the wave's amplitude (Fig. \ref{fig4}b). It is to be noted
that there is no physical wall at the boundaries of the waveguide, and
the elastic sea-bed is only counteracting the spreading of waves which
allows long distance propagation and transmission of wave energy.

\begin{figure}[!h]
  \centering  
  \includegraphics[width=3.3in]{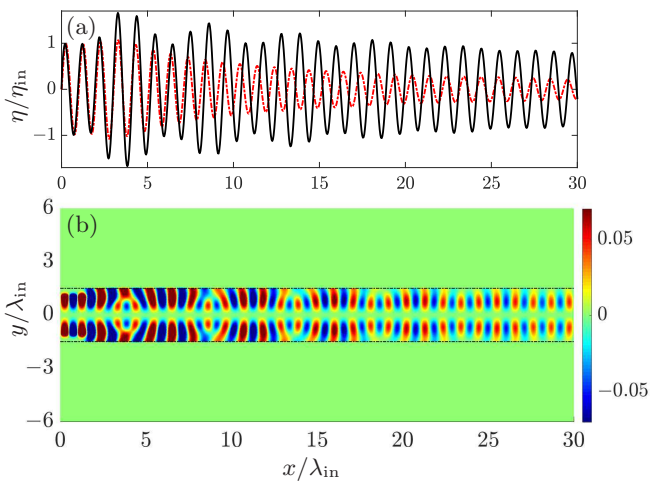}
  \put(-222,157){(a)}
  \put(-245,42){\rotatebox{90}{$y/\lambda_{\text{in}}$}}
  \put(-245,123){\rotatebox{90}{$\eta/\eta_{\text{in}}$}}
  \put(-222,82){(b)}
  \put(-140,-10){$x/\lambda_{\text{in}}$}
  \caption{(a) Wave amplitude along the center-line of numerical
    domain $y=0$ for the case with designed sea-bed carpet ($\L$) and flat
    sea-bed ($\red{\dashL}$). (b) Normalized elevation of the elastic sea-bed
    carpet $\zeta/\eta_{\text{in}}$. The elastic sea-bed domain is $[0,30]\times [-1.5,1.5]$
    over which the elevation is nonzero. The maximum elevation
    recorded over the sea-bed is
    $\zeta_{\max}/\eta_{\text{in}} = 0.07$.}\label{fig5}
\end{figure}

In order to quantify the performance of the waveguide, the surface
wave amplitude along the center line of the waveguide $y=0$ is plotted
in Fig. \ref{fig5}a.  The wave amplitude at the centerline of the
waveguide when the bottom topography is flat has $80\%$ reduction due
to spreading ($\eta_{\text{out}}/\eta_{\text{in}} = 0.2$) as it
travels the distance of the waveguide $D = 30\lambda_{\text{in}}$
where $\lambda_{\text{in}}$ is the incident wavelength. On the
contrary, waves inside the graded refractive index elastic sea-bed
waveguide has only an amplitude decrease of $20\%$ reduction with
$\eta_{\text{out}}/\eta_{\text{in}} = 0.8$ (Fig. \ref{fig5}a).
Running the simulation for an even larger domain,
$D/\lambda_{in} = 35$, we observe a $90\%$ reduction for the case of
flat topography and $25\%$ reduction for the waveguide with elastic
seabed. In order to capture the motion of elastic sea-bed, the elastic
sea-bed elevation is plotted in Fig. \ref{fig5}b. As shown in this
figure, the amplitude of the elastic carpet is at most $7\%$ of the
incident wave amplitude, which makes it suitable for practical
applications. Note that increasing the value of $\delta$ as shown in
Fig. \ref{fig3}, would increase the values of effective gravity at the
boundaries. This would decrease the amount of energy escaped from the
boundaries of the waveguide and therefore increase the final magnitude
of the surface elevation at the final point.

\section{Conclusion}

In this manuscript, we showed that the net effect of a visco-elastic
sea-bed can be summarized through an effective (modified)
gravitational acceleration term which is adjustable through elasticity
and viscosity of the sea-bed parameters. This method proposes a
generic broadband mechanism for modifying the effective gravitational
acceleration which enables new ways of controlling oceanic wave
energy. As an example, we designed a passive elastic sea-bed which is
capable of guiding and transmitting water waves over long
distances. The proposed method has potential applications in guiding
and steering waves toward or away from destinations of interest such
as artificial surf zones, wave energy farms, or to protect shorelines.

%

\end{document}